\documentclass[oldversion]{aa}  
\usepackage{graphicx,amsmath}
\usepackage{color}
\usepackage{natbib}	
\usepackage{url}
\bibpunct{(}{)}{;}{a}{}{,} 

\begin{document}

   \title{Atmospheric escape from HD\,189733b observed in H\,{\sc i} Lyman-$\alpha$: \\ \smallskip{}  
   detailed analysis of HST/STIS September 2011 observations}

   \author{
   V.~Bourrier\inst{1,2}
   A.~Lecavelier des Etangs\inst{1,2}
 \and
H.~Dupuy\inst{1,2} 
 \and
D. Ehrenreich\inst{3}			
\and
A.~Vidal-Madjar\inst{1,2} 
\and
G.~H\'ebrard\inst{1,2} 
\and
G.~E.~Ballester\inst{4}   
\and
J.-M.~D\'esert\inst{5} 	    
\and
R.~Ferlet\inst{1,2}         
\and
D.~K.~Sing\inst{6}     
 \and
P.J. Wheatley\inst{7}         
 }
   
\authorrunning{V.\ Bourrier et al.}
\titlerunning{Atmospheric escape observed in H\,{\sc i} Lyman-$\alpha$}

\offprints{V.B. (\email{bourrier@iap.fr})}

   \institute{
   CNRS, UMR 7095, 
   Institut d'astrophysique de Paris, 
   98$^{\rm bis}$ boulevard Arago, F-75014 Paris, France
   \and
   UPMC Univ. Paris 6, UMR 7095, 
   Institut d'Astrophysique de Paris, 
   98$^{\rm bis}$ boulevard Arago, F-75014 Paris, France
   \and
   Observatoire astronomique de l'Universit\'e de Gen\`eve, 51 Chemin des Maillettes, 1290 Sauverny, Switzerland
   \and
   Lunar and Planetary Laboratory, University of Arizona, 1541 E. University Blvd., Tucson, AZ 85721-0063, USA
   \and   
   Harvard-Smithsonian Center for Astrophysics, 60 Garden Street, Cambridge, MA 02138
   \and
	 Astrophysics Group, School of Physics, University of Exeter, Stocker Road, Exeter EX4 4QL, UK
	 \and   
   Department of Physics,University of Warwick, Coventry CV4 7AL, UK
   }
   
   \date{} %Received ...; accepted ...}
 
  \abstract
{

Observations of transits of the hot giant exoplanet \object{HD\,189733b} in the unresolved H\,{\sc i} Lyman-$\alpha$ line show signs of hydrogen escaping the upper atmosphere of the planet. New resolved Lyman-$\alpha$ observations obtained with the STIS spectrograph onboard the Hubble Space Telescope in April 2010 and September 2011 confirmed that the planet is evaporating, and furthermore discovered significant temporal variations in the physical conditions of its evaporating atmosphere. Here we present a detailed analysis of the September 2011 observations of HD\,189733b, when an atmospheric signature was detected. We present specific methods to find and characterize this absorption signature of escaping hydrogen in the Lyman-$\alpha$ line, and to calculate its false-positive probability, found to be 3.6\%. Taking advantage of the spectral resolution and high sensitivity of the STIS spectrograph, we also present new results on temporal and spectro-temporal variability of this absorption feature. We also report the observation of HD\,189733b in other lines (Si\,{\sc iii} at 1206.5\,\AA, N\,{\sc v} at 1240\,\AA). Variations in these lines could be explained either by early occultation by a bow-shock rich in highly ionized species, or by stellar variations. 

}

\keywords{planetary systems - Stars: individual: HD\,189733}

   \maketitle

\section{Introduction and observations}

\subsection{Evaporation of hot Jupiters}

Although the phenomenon of evaporation had been anticipated by \citet{Burrows1995} and \citet{Guillot1996}, the existence of a large number of hot Jupiters surviving atmospheric escape suggested that evaporation processes should be modest. In this frame it came as a surprise to discover that the exoplanet HD\,209458b was losing gas (\citealt{VM2003}). Transit observations of the Lyman-$\alpha$ line showed excess absorption due to an extended cloud of neutral hydrogen H\,{\sc i}  and constrained the (H\,{\sc i}) escape rate with a lower limit of $10^{10}$\,g\,s$^{-1}$ (\citealt{VM2003}; \citealt{VM_lecav2004}). This discovery has been challenged by \citet{BJ2007}, but the apparent discrepancy has been resolved and the result obtained by Ben-Jaffel on this first data set strengthens the evaporation scenario (\citealt{VM2004}; \citealt{VM2008}). It was also confirmed by two subsequent observations at low spectral resolution: a 5\% absorption of the whole Lyman-$\alpha$ line was indeed measured using unresolved emission line flux during planetary transits observed with the Space Telescope Imaging Spectrograph (STIS) instrument (\citealt{VM2004}) and the Advanced Camera for Surveys (ACS) instrument (\citealt{Ehrenreich2008}) onboard the Hubble Space Telescope (HST). An independent analysis of the low-resolution data set used by \citet{VM2004} has confirmed that the transit depth in Lyman-$\alpha$ is significantly greater than the transit depth due to the planetary disk alone (\citealt{BJ_Hosseini2010}). These three independent observations show a significant amount of gas at velocities exceeding the planet escape velocity, leading to the conclusion that HD\,209458b is evaporating. This conclusion is strengthened by the observations of absorption in the O\,{\sc i} and C\,{\sc ii} lines (\citealt{VM2004}), which led to the identification of a blow-off escape mechanism. This is also confirmed by recent observations of absorption in the C\,{\sc ii} and Si\,{\sc iii} lines using the HST Cosmic Origins Spectrograph (\citealt{Linsky2010}). Note that additional analyses of these observations are being made (Ballester \& Ben-Jaffel, in prep.). A similar conclusion has been reached in the case of WASP-12b with observations of Mg\,{\sc ii} and near UV broad-band transit absorption (\citealt{Fossati2010}).

Based on these observational constraints, several theoretical models have been developed to explain and characterize the evaporation processes (\citealt{Lammer2003}; \citealt{Lecav2004}, \citeyear{Lecav2007}, \citeyear{Lecav2008b}; \citealt{Baraffe2004,Baraffe2005,Baraffe2006}; \citealt{Yelle2004,Yelle2006}; \citealt{Jaritz2005}; \citealt{Tian2005}; \citealt{GarciaMunoz2007}; \citealt{Holmstrom2008}; \citealt{Stone2009}; \citealt{MurrayClay2009}; \citealt{Adams2011}; \citealt{Guo2011}). In the models developed by \citet{Holmstrom2008} and \citet{Ekenback2010}, the observed hydrogen is produced through charge exchange between the stellar wind and the planetary escaping exosphere. Even in these last models, neutral hydrogen escaping the planet is required for charge exchange with hot protons of the stellar wind. Therefore, all models led to the conclusion that most of the EUV and X-ray energy input by the host star is used by the atmosphere to escape the planetary gravitational potential (\citealt{Owen2012}; for detailed estimates of the EUV and X-ray input, see \citealt{Sanz-Forcada2011}). With this idea in mind, \citet{Lecav2007} developed an energy diagram in which the potential energy of the planet is plotted versus the stellar energy received by its upper atmosphere. For HD\,209458b, the escape rates derived from this diagram are not high enough for the evaporating exoplanet to lose a significant amount of its mass (\citealt{Ehrenreich_desert2011}). However, the nature of planets with smaller masses and closer orbits could be significantly altered by evaporation, and these planets would end as planetary remnants (\citealt{Lecav2004}). The recently discovered CoRoT-7b (\citealt{Leger2009}) and Kepler-10b (\citealt{Batalha2011}) could be examples of a new category of planetary remnants, which were proposed for classification as ``chthonian planets'' (see \citealt{Lecav2004}).

For a few years, the only observed evaporating planet was HD\,209458b, and many questions remained to be answered. What is the evaporation state of other hot Jupiters and very hot Jupiters? Is evaporation a common process amongst hot Jupiters? How does the planetary system and stellar characteristics influence the escape rate? Some light has been shed on these questions thanks to the discovery of HD\,189733b (\citealt{Bouchy2005}), a planet transiting a bright and nearby K star (V=7.7). Among the stars harboring transiting planets, HD\,189733b belongs indeed to the second brightest at Lyman-$\alpha$. Despite the failure of the HST/STIS instrument one year before the discovery of HD\,189733b, \citet{Lecav2010} detected atmospheric escape through Lyman-$\alpha$ transit observations with the HST/ACS instrument. Moreover, \citet{Lecav2012} compared two transit observations of HD\,189733b made in April 2010 and September 2011 with the UV channel of the repaired HST/STIS (using the same methodology as employed by \citealt{VM2003} for HD\,209458b), and showed that the escape from this planet is subject to significant temporal variations. In this paper we present a detailed analysis of the STIS transit observations of HD\,189733b made in September 2011. We also show a new time-resolved spectral signature of the absorption observed at this epoch by using time-tagged photon count data. We also present the transit observed in other lines, such as Si\,{\sc iii} at 1206.5\,\AA\ and N\,{\sc v} at 1239\,\AA\ and 1243\,\AA.

\subsection{HD\,189733b}
  
The very hot Jupiter HD\,189733b is located 19.3~parsecs away from Earth, with a semi-major axis of 0.03~AU and an orbital period of 2.2~days. The planetary transits and eclipses can be used to probe its atmosphere and environment ({\it e.g.}, \citealt{Charbonneau2008}; \citealt{Desert2009}).   
HD\,189733b orbits a bright main-sequence star, with a magnitude V=7.7. This K2V star emits one of the highest Lyman-$\alpha$ flux ever measured for transits observations (with the noticeable exception of 55 Cnc; \citealt{Ehrenreich2012}), which makes HD\,189733b a particularly good candidate to study atmospheric evaporation. 

Thanks to the proximity of HD\,189733, and the large surface covered by the planet against its relatively small star, extensive observational studies have been made of this hot Jupiter. The planet has a mass $M_p$=1.13~Jupiter masses ($M_{\rm Jup}$) and a radius $R_p$=1.16~Jupiter radii ($R_{\rm Jup}$) in the visible (\citealt{Bakos2006}; \citealt{Winn2007}). The planetary disk transit results in a $\approx$2.4\% occultation depth from visible to near infrared (\citealt{Desert2009}; \citealt{Sing2011}). The short period of the planet (2.21858~days) has been measured precisely (\citealt{Hebrard2006}; \citealt{Knutson2009}). Spectropolarimetry has measured the strength and topology of the {\it stellar} magnetic field, which reaches up to 40~G (\citealt{Moutou2007}; \citealt{Fares2010}). Sodium has been detected in the planet atmosphere by both ground-based (\citealt{Redfield2008}) and space-based observations (\citealt{Huitson2012}). Using the HST/ACS, \citet{Pont2008} detected atmospheric haze, which is interpreted as Mie scattering by small particles (\citealt{Lecav2008a}). High signal-to-noise HST/NICMOS/STIS observations (\citealt{Sing2009,Sing2011}) and HST/WFC3 (\citealt{Gibson2012}) have shown that the near-IR spectrum below 2~$\mu$m is due to haze scattering and/or water absorption (\citealt{Swain2008}; see also \citealt{Gibson2011}), and Rayleigh scattering might even extend to longer wavelengths (\citealt{Pont2012}). The detection of an H$_2$O signature using transit photometry (\citealt{Tinetti2007}) has been subjected to controversy (\citealt{Ehrenreich2007}; \citealt{Agol2010}; \citealt{Desert2009, Desert2011}). It has been tentatively proposed that CO molecules can explain the excess absorption seen at 4.5~$\mu$m (\citealt{Charbonneau2008}; \citealt{Desert2009}; \citealt{Knutson2012}). Applying best-estimate approaches to dayside infrared emission spectra of HD\,189733b, \citet{Lee2012} and \citet{Line2012} both reported the detection of H$_2$O and CO$_2$ in the atmosphere (see also \citealt{Swain2009}). Using Spitzer spectroscopy of planetary eclipses, \citet{Grillmair2008} also found evidence of H$_2$O absorption signatures and possibly weather-like variations in the atmospheric conditions. Note that \citet{Agol2010} found an upper limit of 2.7\% on the variability of the dayside planet flux, which rules out the most extreme weather fluctuations on HD\,189733b. An alternative source for the observed variability may be the change of location of a dayside hot spot detected at an offset from the substellar point (\citealt{Knutson2007}, \citealt{Agol2010}, \citealt{Majeau2012}). A sensitive search with GMRT has provided very low upper limits on the meter-wavelength radio emission from the planet, indicating a weak planetary magnetic field (\citealt{Lecav2009,Lecav2011}).

\subsection{Previous Lyman-$\alpha$ observations of HD\,189733b}
\label{previous obs}

In 2007-2008, because of the failure of the Space Telescope Imaging Spectrograph (HST/STIS) in 2004, HD\,189733 was observed with the Solar Blind Camera of the (HST/ACS). Transit observations showed a diminution of 5.05$\pm$0.75\% of the entire H\,{\sc i} Lyman-$\alpha$ curve (\citealt{Lecav2010}). This was more than the 2.4\% planetary disk occultation depth and was interpreted as the result of the evaporation of the planet's upper atmosphere. However, the limited spectral resolution of the ACS data in the far-ultraviolet wavelengths of the \mbox{Lyman-$\alpha$} line (at 1215.67\,\AA ) implies that no conclusion could be made on the radial velocities of the escaping gas, and this detection called for more observations. After the refurbishment of HST in May 2009, HD\,189733b was thus observed during transit in April 2010 with the G140M grating of the repaired HST/STIS. Disappointingly, no significant absorption was detected in the Lyman-$\alpha$ line (2.9$\pm$1.4\%, see \citealt{Lecav2012}), other than the occultation depth by the planetary disk alone. \\
Note that \citet{Jensen2012} reported a significant absorption feature at H$\alpha$ in the transmission spectra of HD\,189733b. This transit-dependent absorption was detected within a narrow band at the line center (-24.1 to 26.6\,km\,s$^{-1}$). Because of the limited resolution of the ACS spectra, Jensen et al. could not directly compare the velocity ranges between their H-$\alpha$ absorption detection and the H\,{\sc i} Lyman-$\alpha$ absorption feature reported by \citet{Lecav2010}. Although the stellar Lyman-$\alpha$ emission line is strongly absorbed by the interstellar hydrogen over their observed velocity range, our recent and highly resolved STIS observations could enable a more direct comparison between the two atmospheric signatures.

\section{Observations}
The observations of HD\,189733b in the Lyman-$\alpha$ line performed in 2007 - 2008 showed significant absorption in the exosphere of the planet, while in April 2010 no atmospheric absorption signature was detected. To address the question of these variations, new transit observations of the planet were made in September 2011 with the G140M grating of the HST/STIS. The log of these time-tagged observations is given in Table~\ref{Obs Log}. The analysis described in this paper was made with 1D spectra extracted using CALSTIS (version 3.32) data pipeline. These spectra cover the far-UV wavelengths from 1195 to 1248\,\AA\ with a spectral resolution of about 20\,km\,s$^{-1}$ at 1215.67\,\AA. The planet was observed during four consecutive orbits of the HST around the Earth: two orbits before the transit, one during the transit, and one after the transit. The first exposure lasts 1800\,s and the three others about 2100\,s. Data acquisitions are interrupted by Earth occultation lasting about 3500\,s. Contrary to ACS observations, for which the co-addition of three independent transits was necessary to detect the transit signature in the Lyman-$\alpha$ line, the highest sensitivity of the STIS data allows an independent study of each transit observation.

\begin{table*}[tbh]
\begin{tabular}{clcccc}
\hline
\hline
\noalign{\smallskip}
Data set & Date & \multicolumn{2}{c}{Observation} & \multicolumn{2}{c}{Time from center of transit}   \\
         &      & Start & End 										& Start & End			   \\
\noalign{\smallskip}
\hline
\noalign{\smallskip}
HST orbit \#1 & 2011-09-07     & 19:49:21 & 20:20:55  & -03:31:02   & -02:59:28 \\
HST orbit \#2 & 2011-09-07     & 21:18:37 & 21:56:10  & -02:01:46   & -01:24:13 \\
HST orbit \#3 & 2011-09-07     & 22:54:26 & 23:31:59  & -00:25:57   &  \ 00:11:35 \\
HST orbit \#4 & 2011-09-08     & 00:30:16 & 01:07:49  &  \ 01:09:52 &  \ 01:47:25 \\
\noalign{\smallskip}
\hline
\hline
\end{tabular}
\caption{Log of the September 2011 observations. Time is given in UT.}
\label{Obs Log}
\end{table*}

\section{Data analysis}
\label{data ana}

\subsection{Resulting spectra}
\begin{figure}[tbh]
\includegraphics[width=\columnwidth]{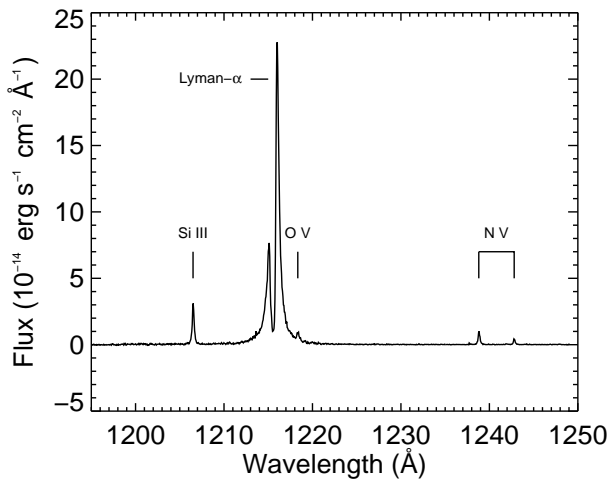}
\caption[]{Plot of the wavelength-calibrated STIS spectrum of HD\,189733b with the Lyman-$\alpha$, Si\,{\sc iii}, O\,{\sc v}, and N\,{\sc v} stellar emission lines.} 
\label{stellarlines}
\end{figure}

We identified in the STIS spectra the stellar emission lines of Si\,{\sc iii} (1206.5\,\AA), O\,{\sc v} (1218.3\,\AA), the N\,{\sc v} doublet (1242.8\,\AA\ and 1238.8\,\AA), and the bright H\,{\sc i} Lyman-$\alpha$ line (1215.67\,\AA) (Fig.~\ref{stellarlines}). To calculate the radial velocities relative to the star, we need to estimate the radial velocity of the star relative to the STIS wavelength calibration. Using the stellar emission lines we found that the star has a red-shifted velocity of $\sim$5\,km\,s$^{-1}$ (heliocentric).

In order to detect possible transit signatures of HD\,189733b and its atmosphere, we calculated for each stellar line the time evolution of the total emission flux, or of the flux within a given wavelength range. We also compared the April 2010 and September 2011 observations of the stellar lines other than the Lyman-$\alpha$ line (for comparison of this line between 2010 and 2011, see \citealt{Lecav2012}). A signature of the atmosphere is considered to be detected when an excess absorption is seen in addition to the absorption by the planetary disk itself during the planet transit.

\subsection{N\,{\sc v} doublet (at 1238.8\,\AA) and O\,{\sc v} line}
For the April 2010 and September 2011 observations, we compared the total flux in the stellar Si\,{\sc iii} emission line and the N\,{\sc v} doublet before, during, and after the transit. Results are summarized in Table ~\ref{SIII Log}. We do not detect any variation showing a transit-like signal in the 2010 or in the 2011 observations for the N\,{\sc v} line at 1238.8\,\AA. All variations, in particular differences in flux between 2010 and 2011, are most likely attributable to stellar variations in a line arising from an active region of the stellar atmosphere. The analysis of the O\,{\sc v} line, fainter and located within the Lyman-$\alpha$ line red wing, gave no significant results.

\subsection{Si\,{\sc iii} line and N\,{\sc v} doublet (at 1242.8\,\AA)}

\begin{figure}[tbh]
\begin{center}
\includegraphics[angle=-90,scale=.33]{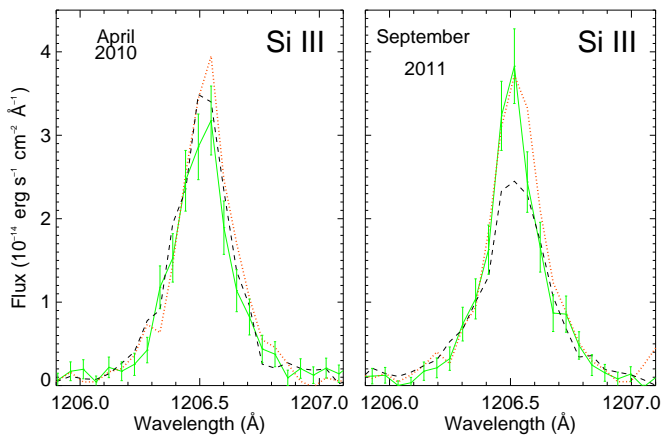}
%\medskip
\end{center}
\caption[]{Plot of the flux in the Si\,{\sc iii} line for the 2010 (left panel) and 2011 (right panel) observations. The black dashed line shows the spectrum before the planet transit; the green solid line with error bars at the 1\,$\sigma$ level shows the flux during the transit; the red dotted line shows the flux after the transit.} 
\label{SiIII line}
\end{figure}

\begin{figure}[tbh]
\begin{center}
\includegraphics[angle=-90,scale=.33]{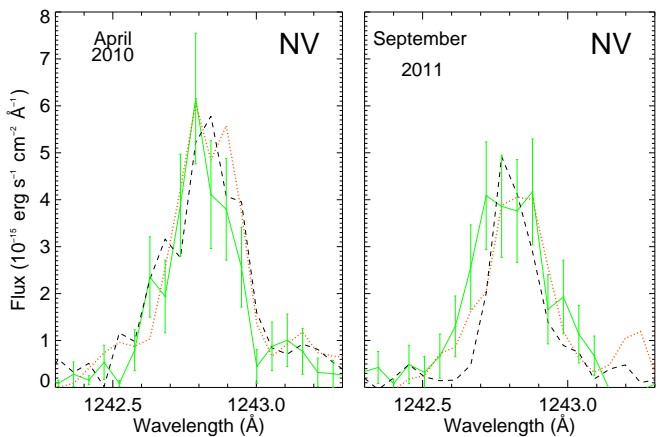}
%\medskip
\end{center}
\caption[]{Plot of the flux in the N\,{\sc v} line between 1242.5 and 1243.1\,\AA, for the 2010 (left panel) and 2011 (right panel) observations.  The black dashed line shows the spectrum before the planet transit; the green solid line with error bars at the 1\,$\sigma$ level shows the flux during the transit; the red dotted line shows the flux after the transit.} 
\label{Nv line}
\end{figure}

Although there is no significant variation in the first epoch observations for the Si\,{\sc iii} line and the N\,{\sc v} doublet, in the second epoch we see significant time variations of the flux measured in the Si\,{\sc iii} line, and in the N\,{\sc v} line at 1242.8\,\AA\ (see Table ~\ref{SIII Log}). In the 2011 observations, these stellar lines are observed to be fainter before the transit than during and after the transit (Fig ~\ref{SiIII line} and Fig ~\ref{Nv line}). There is a 3.5\% probability to find the high variations observed between the fluxes before, during, and after the transit for the N\,{\sc v} line at 1242.8\,\AA\ (see the $\chi^2$ values in Table ~\ref{SIII Log}). This probability is 0.6\% for the Si\,{\sc iii} line. Variations in the Si\,{\sc iii} line brightness have already been observed in the planetary system of HD\,209458b (\citealt{Linsky2010}). However, here we see a higher flux during the transit, which cannot be interpreted by a classical absorption in the planetary exosphere. Our observation could be interpreted by an early ingress caused by the formation of a bow-shock surrounding the planet magnetosphere. This was modeled by \citet{Llama2011} and \citet{Vidotto2010,Vidotto2011a,Vidotto2011b}, and explains for example the WASP-12b observations of an early ingress in the Mg\,{\sc ii} line (\citealt{Fossati2010}, \citealt{Haswell2012}). Alternatively, intrinsic stellar line variations could be the explanation of our observations. This raises the question of possible stellar line variations triggered by the planet, as already suspected in the case of the C\,{\sc ii} flare detected in ACS observations of the same planet (\citealt{Lecav2010}). The Si\,{\sc iii} line and N\,{\sc v} doublet thus appear as possible candidates to diagnose the star-planet interaction (SPI), and additional observations are needed to address this possibility. Hereafter we consider only the Lyman-$\alpha$ observations.

\begin{table*}[tbh]
\begin{tabular}{lcccccc}
\hline
\hline
\noalign{\smallskip}
					& \multicolumn{2}{c}{Si\,{\sc iii} line} 										 &     \multicolumn{4}{c}{ N\,{\sc v} doublet} \\
					& \multicolumn{2}{c}{Flux within 1206 - 1207\,\AA}	   			 & \multicolumn{2}{c}{Flux within 1238.4 - 1239.1\,\AA}							& \multicolumn{2}{c}{Flux within 1242.5 - 1243.1\,\AA}	   			 \\
					&	\multicolumn{2}{c}{(10$^{-14}$\,erg\,s$^{-1}$\,cm$^{-2}$)} &	\multicolumn{2}{c}{(10$^{-15}$\,erg\,s$^{-1}$\,cm$^{-2}$)} 			&	\multicolumn{2}{c}{(10$^{-15}$\,erg\,s$^{-1}$\,cm$^{-2}$)}  					 \\			
					&	 2010 & 2011	 																						 &	 2010 & 2011																&	 2010 & 2011	 \\	
\noalign{\smallskip}
\hline
\noalign{\smallskip}
Before transit & 1.05$\pm$0.04     & 0.87$\pm$0.04  & 3.59$\pm$0.19 &	2.77$\pm$0.16	& 1.74$\pm$0.13	& 0.97$\pm$0.10  \\
During transit & 0.94$\pm$0.05     & 0.98$\pm$0.05  & 3.15$\pm$0.23 &	2.72$\pm$0.22	& 1.50$\pm$0.16	& 1.40$\pm$0.15  \\
After transit  & 1.07$\pm$0.06     & 1.08$\pm$0.06  & 3.07$\pm$0.23 &	2.77$\pm$0.22	& 1.75$\pm$0.17	& 1.19$\pm$0.14  \\
\noalign{\smallskip}
\hline
\noalign{\smallskip}
$\chi^2$       & 3.82								&10.1					&3.86							&0.04						&1.65						&6.70							\\
\noalign{\smallskip}
\hline
\hline
\end{tabular}
\caption{Total flux in the Si\,{\sc iii} emission line (at 1206.5\,\AA) and the N\,{\sc v} emission line (at 1242.8\,\AA\ and 1238.8\,\AA) before, during, and after the transit. The $\chi^2$ values are the standard deviations associated to each stellar line in either 2010 or 2011.}
\label{SIII Log}
\end{table*}

\subsection{Lyman-$\alpha$ line}

\subsubsection{The stellar Lyman-$\alpha$ emission line profile}

\begin{figure}[tbh]
\includegraphics[width=\columnwidth]{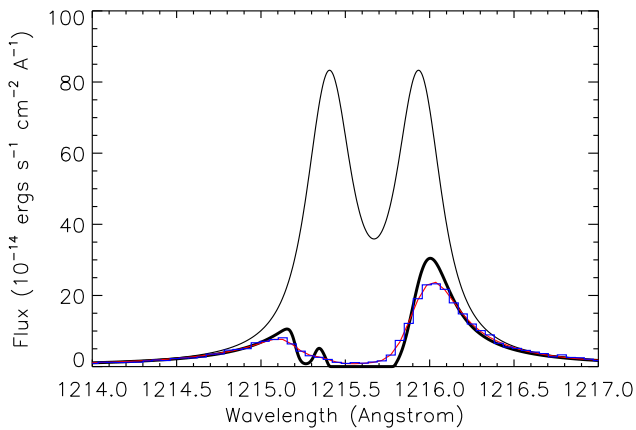}
\caption[]{Plot of the theoretical profile of \object{HD\,189733} Lyman-$\alpha$ line. The black thin line shows the theoretical intrinsic stellar emission line profile as seen by hydrogen atoms escaping the planetary atmosphere. The black thick line shows the resulting profile after absorption by the interstellar hydrogen  (1215.6\AA) and deuterium (1215.3\AA). The line profile convolved with the HST G140M instrumental line spread function (red line) is compared to the observations (blue histogram), yielding a good fit with a $\chi^2$ of 36.1 for 40 degrees of freedom.} 
\label{fit theo spectrum}
\end{figure}

The Lyman-$\alpha$ line is the brightest stellar line in the STIS spectra from 1195 to 1248\,\AA. The single stellar Lyman-$\alpha$ line is massively absorbed by the interstellar atomic hydrogen H\,{\sc i} in a narrow band at the line center (1215.6\,\AA), and to a lesser extent by the interstellar deuterium D\,{\sc i} around 1215.3\,\AA. As a result, the spectrally resolved line, at the resolution of the G140M spectrograph, is composed of two peaks separated by this deep absorption. In order to estimate the Lyman-$\alpha$ line flux and profile, we fitted the line profile as observed in September 2011, following the method of \citet{Wood2005} and \citet{Ehrenreich2011}. The best result was obtained by modeling the stellar emission line using two Voigt profiles with the same width and damping constant. The free parameters of the fit are the total stellar flux, the Voigt parameters (width, damping constant, and the wavelength difference between the two profiles centers), and the ISM parameters (H\,{\sc i} column density, ISM and star radial velocities, and turbulence velocity). We used a D/H ratio of 1.5$\times$10$^{-5}$\ ({\it e.g.}, \citealt{Hebrard_Moos2003}, \citealt{Linsky2006}). Finally we added a single free parameter to fit the strength of the wing of the line spread function (LSF), which is added to the G140M published tabulated LSF\footnote{\url{http://www.stsci.edu/hst/stis/performance/spectral_resolution}}. The wing of this LSF is taken as a Gaussian with a fixed width measured in the published G140M LSF to be $\sigma_{Wing\,LSF}=4.6$\,pixels. The comparison between the best model reconstructed Lyman-$\alpha$ stellar profile and the observed spectrum yields a $\chi^2$ of 36.1 for 49 data points in the wavelength range 1214.25-1215.5\,\AA\ and 1215.8-1217.1\,\AA\ (see Fig.~\ref{fit theo spectrum}). The resulting profile shows a double peak emission, which is usually seen in the Lyman-$\alpha$ line of cool stars, and can be interpreted as self-absorption in the chromosphere of the star (\citealt{Wood2005}).

In the search for the best model to fit the Lyman-$\alpha$ profile, we used the Bayesian Information Criterion (BIC) and the Akaike Information Criterion (AIC) (see, e.g., \citealt{deWit2012}). These criteria prevent us from over-fitting and are based on the likelihood function given by the $\chi^2$ and on a penalty term related to the number of parameters in the fitting model (\citealt{Crossfield2012}; \citealt{Cowan2012}). The lowest BIC is obtained for the reference model described above (see parameters in Table~\ref{Table:BIC}). When we allowed for the temperature of the ISM to be a free parameter (allowing different turbulent widths for the ISM hydrogen and deuterium absorption lines), we obtained a better $\chi^2$ but the larger BIC and AIC show that this decrease cannot be interpreted in terms of the information content of the data. We also obtained a better fit to the data by using a two-Gaussian LSF with three additional free parameters, but again the larger BIC and AIC shows that the increase of freedom is responsible for the apparent improvement in the fit. We also concluded that the better fit to the profile obtained by using two different Voigt profiles (with different widths and damping constants) is not significant. We also tested if more simple emission line profiles with lower degrees of freedom could be used (one single-Voigt profile or two-Gaussian profile) and concluded that the reference model with two similar Voigt profiles provides the best fit to the data.

\begin{table}
\begin{tabular}{lrrrrr}
\hline
\noalign{\smallskip}
Model & $\chi ^2$ & k & DOF & BIC & AIC \\
\noalign{\smallskip}
\hline
Reference model 		    &  36.1 &  9 & 40 &  71.1 &  54.1 \\
$T_{\rm ISM} \ne 0$         &  35.8 & 10 & 39 &  74.7 &  55.8 \\
Double Gaussian LSF     &  34.4 & 12 & 37 &  81.1 &  58.4 \\
2 different Voigt profiles    & 35.5 &  11  & 38 &   78.3 &  57.5 \\
1-Voigt profile    &  41.5 &  8 & 41 &  72.6 &  57.5 \\
2-Gaussian profile & 166.6 &  8 & 41 & 197.7 & 182.6 \\
\noalign{\smallskip}
\hline
\end{tabular}
\caption{$\chi^2$, BIC and AIC for various models with $k$ degrees of freedom used to fit the Lyman-$\alpha$ line profile.
\label{Table:BIC}}
\end{table}

\subsubsection{Geo-coronal emission}

\begin{figure}[tbh]
\includegraphics[angle=-90,width=\columnwidth]{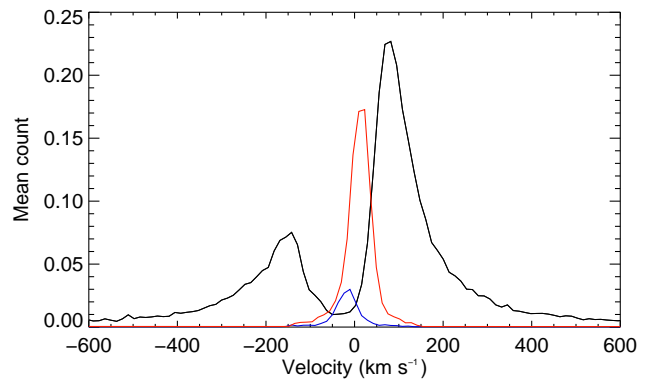}
\caption[]{Plot of HD\,189733b Lyman-$\alpha$ line profile (black line) superimposed with the geo-coronal emission (red and blue lines). The red line corresponds to the 2010 observations, the blue line to the 2011 observations. All lines are averaged over the four HST orbits. The geo-coronal emission is noticeably low during the 2011 observations.} 
\label{airglow}
\end{figure}

In the raw data, the stellar emission line is superimposed with the geo-coronal airglow emission from the upper atmosphere of the Earth (\citealt{VM2003}). This emission line can be well estimated and removed from the final spectrum using the calstis data pipeline (version 2.32 of November 5, 2010). Independent re-analysis of raw data using the same methodology as in \citet{VM2003} and \citet{Desert2004} confirmed that the airglow emission can be efficiently subtracted and included in the final error budget. Moreover, because we used a narrow spectrograph slit of 0.1", the airglow contamination is limited to the central part of the Lyman-$\alpha$ line and does not contaminate the part of the spectrum where atmospheric signatures are detected, in the wings of the line (Sect.~\ref{atm_h}). The airglow emission may be affected by daily and seasonal variations, and the data of September 2011 present a noticeably low airglow emission level (Fig.~\ref{airglow}).

\subsubsection{Telescope "breathing"}
\label{breathing}

\begin{figure}[tbh]
\includegraphics[angle=-90,width=\columnwidth]{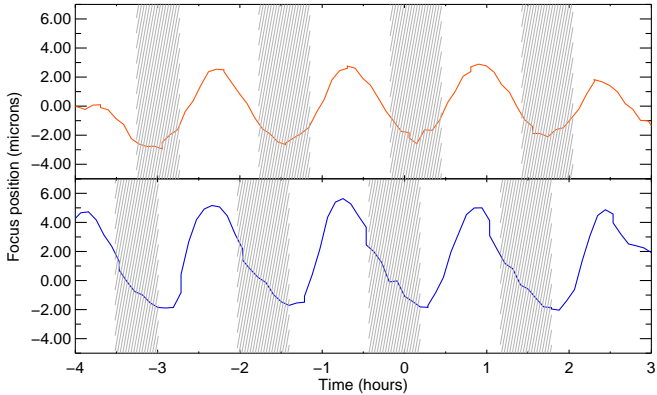}
\caption[]{Longitudinal motion of the secondary mirror for the 2010 observations (top panel) and the 2011 observations (bottom panel). Time is given relative to the center of the planetary transit. Hatched zones show the time windows of our observations of HD\,189733b.} 
\label{defocus}
\end{figure}

Using time-tagged acquisitions of the STIS spectra, we checked variations at short time scales within a given HST orbit in a similar manner as done by \citet{BJ2007}. In the 2011 observations, we see a periodic variation of the measured Lyman-$\alpha$ flux in phase with the HST position on its orbit, with the same amplitude of about 10\% at all wavelengths and no spectral signature. These variations are not seen at a significant level in the 2010 observations. They are interpreted by the STScI STIS experts as caused by the telescope "breathing", which is known to affect STIS observations in the UV and in the visible because of variations of the telescope throughput. The breathing is caused by changes in temperature that the HST experiences during each orbit. This causes small motions of the secondary mirror in the STIS bench, which alters the alignment of the target in the aperture (\citealt{Kimble1998}). It may also change the focus position and the size of the point spread function, affecting the spectrograph throughput through the narrow 0.1" slit. Although the breathing depends on a large number of parameters and cannot be predicted, it can be modeled for a past observation. Using the STScI Focus model (\citealt{Cox_Niemi2011}), we thus assessed the repeatability of the breathing effect on the focus position during the 2011 observations. While the time windows of these observations correspond to significant variations of the focus position, the 2010 observations coincide with times when the focus position is subjected to only low variations (Fig.~\ref{defocus}). The shape and amplitude of these systematic variations can change between visits of the same target, as was found in STIS optical transit observations of HD\,209458 (\citealt{Brown2001}; \citealt{Charbonneau2002}; \citealt{Sing2008a}) and of HD\,189733 (\citealt{Sing2011}; \citealt{Huitson2012}), and in STIS FUV transit observations of 55\,Cnc (\citealt{Ehrenreich2012}). Extensive experience with optical STIS data over the last decade furthermore indicates that the orbit-to-orbit variations within a single visit are both stable and highly repeatable, except for the first orbit, which displays a different trend (see \citealt{Brown2001} and \citealt{Charbonneau2002}). Because of its repeatability, this effect is easy to correct for.

\begin{figure}[tbh]
\includegraphics[angle=-90,width=\columnwidth]{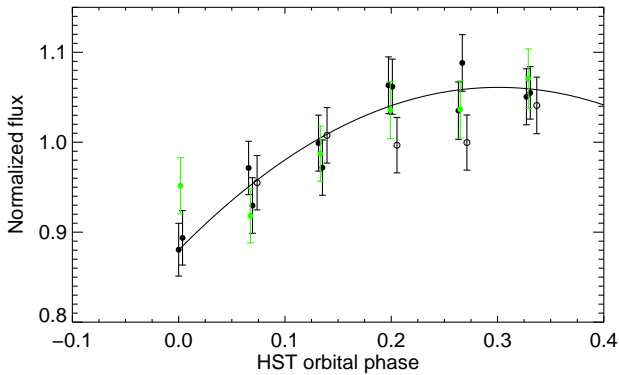}
\caption[]{Lyman-$\alpha$ fluxes for 379~s exposures, phase-folded on the HST orbital period. The empty circles correspond to the first orbit. The green circles correspond to the measurements obtained during the planetary transit. The black line is the second-order polynomial fit to the other orbits measurements (filled black dots). } 
\label{breathingcorrection}
\end{figure}

For each HST orbit, we divided the time-tagged exposures into six 379~s subexposures. Following the method of \citet{Sing2011} and \citet{Brown2001}, we phase-folded these exposures over the HST-orbital period of 96~minutes. Each exposure was then spectrally summed over the entire Lyman-$\alpha$ line, and normalized by the mean value of all the exposures of its related orbit. We excluded the first orbit. We also excluded observations obtained during the planetary transit. We found that the breathing effect was best fitted with a second-order polynomial (Fig.~\ref{breathingcorrection}; see \citealt{Ehrenreich2012} for a similar correction of this effect).

\begin{figure}[tbh]
\includegraphics[angle=-90,width=\columnwidth]{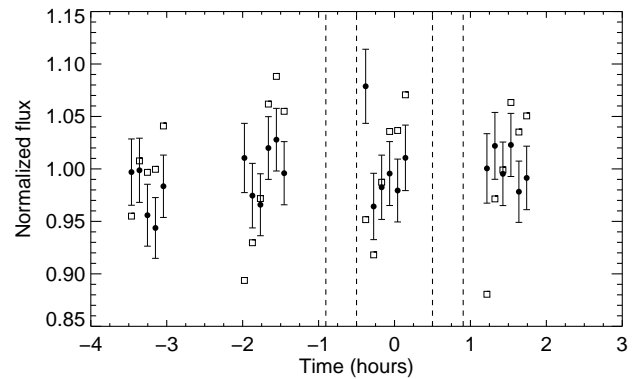}
\caption[]{Raw fluxes not corrected for the breathing effect (empty squares), as a function of time in the September 2011 observations. Time is given relative to the center of the planetary transit. Circles correspond to the corrected flux. Each point stands for a 379~s exposure spectrum, summed over the whole Lyman-$\alpha$ line, and normalized by the flux over the entire corresponding orbit. Vertical dashed lines show the beginning and end of ingress and egress of the transit.} 
\label{breathing_corrected}
\end{figure}

Fig.~\ref{breathingcorrection} shows that the first orbit seems to be lower than the expected unabsorbed Lyman-$\alpha$ stellar light curve. This could mean the absorption signatures described in Sect.~\ref{atm_h}, and obtained with the two orbits before the transit as reference, may be underestimated. 
We correct the data for the breathing effect by dividing the spectra of each 379~s subexposure by the value of the polynomial fit at the same HST orbital phase. This correction does not apply to the first orbit because of its different trend. The spectra are reconstructed for each orbit by coadding their respective exposures (Fig.~\ref{breathing_corrected}). Note that we obtain the same estimates for absorptions and ratios measured using the full HST orbit exposures, with or without the correction, as done by \citet{Lecav2012}. However, the correction must be taken into account for measurements on partial orbit exposures obtained using time-tagged data.

\section{Atmospheric hydrogen}
\label{atm_h}
\subsection{Full Lyman-$\alpha$ flux}

\begin{figure}[tbh]
\includegraphics[angle=-90,width=\columnwidth]{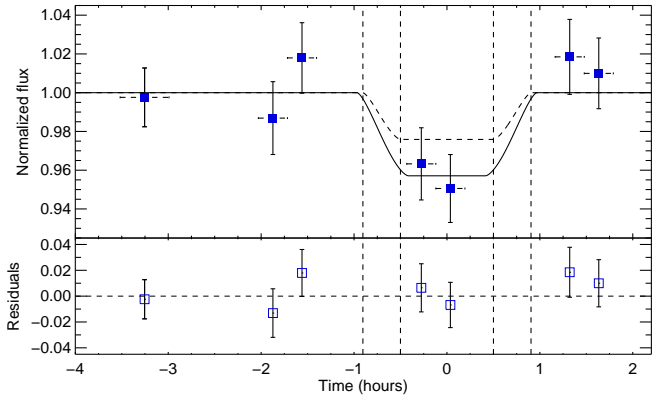}
\caption[]{Plot of the total flux within the whole \mbox{Lyman-$\alpha$} line, as a function of time (blue square symbols). Each HST orbit is divided into two 1127~s independent subexposures, corrected for the breathing effect (except for the first orbit, see Section~\ref{breathing}). Time is given relative to the center of the planetary transit. Vertical dashed lines show the beginning and end of ingress and egress of the transit. Horizontal error bars centered on the symbols show the duration of the exposures. A light curve fitted with a classical planetary occultation model is displayed as a solid black line, with a transit depth in excess compared to the transit depth observed at optical wavelengths (dashed black line). The bottom panel shows the corresponding residuals (empty symbols).}
\label{lyman_4obs_2011}
\end{figure}

In a first step, to search for the transit signature of the planetary atmosphere in the Lyman-$\alpha$ line, we compared the total line flux measured during the transit with the flux measured before the transit. The transit depth by the planetary disk alone as seen from the visible to the near-infrared amounts to 2.4\% (\citealt{Pont2007}; \citealt{Sing2009}; \citealt{Desert2009}). The observations of HD\,189733b made in April 2010 showed no significant excess in the transit absorption depth, with a depth for the flux within the whole Lyman-$\alpha$ line of 2.9$\pm$1.4\%, including the planetary disk occultation (\citealt{Lecav2012}). In contrast, a transit absorption of 5.0$\pm$1.3\% is found in the total flux of the Lyman-$\alpha$ line for the 2011 observations. We fitted the flux measured during the transit with a classical planetary occultation model from \citet{Mandel2002}. The best fit is consistent with the presence of an extended absorbing atmosphere surrounding the planet (see Fig.~\ref{lyman_4obs_2011}). Substracting the planetary disk absorption, the signature measured during the 2011 transit corresponds to an excess absorption, due to atmosperic hydrogen only, of 2.3$\pm$1.4\%. This is consistent with the excess absorption of 2.71$\pm$0.75\% due to hydrogen only, obtained with the HST/ACS observations in 2007-2008 (\citealt{Lecav2010}).

\subsection{Spectrally resolved absorption features}

\begin{figure}[tbh]
\includegraphics[angle=-90,width=\columnwidth]{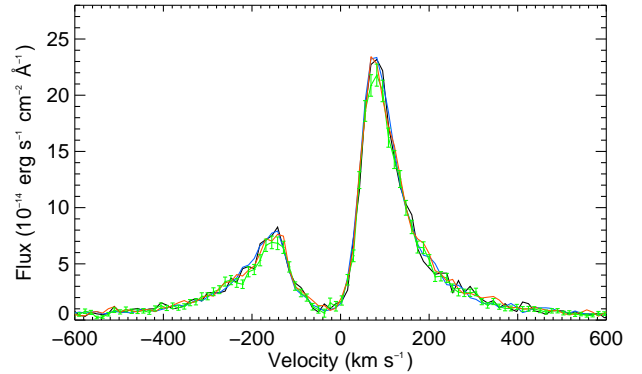}
\caption[]{Lyman-$\alpha$ line profile. Four spectra, corresponding to the four HST orbits of the September 2011 observations, are displayed as a function of radial velocity relative to the star. The black and blue lines show the fluxes before the planet transit; the green line with error bars at the 1\,$\sigma$ level shows the flux during the transit; the red line shows the flux after the transit.} 
\label{Lyman_2011_4orbites}
\end{figure}

\begin{figure}[tbh]
\begin{center}
\includegraphics[angle=-90,scale=.33]{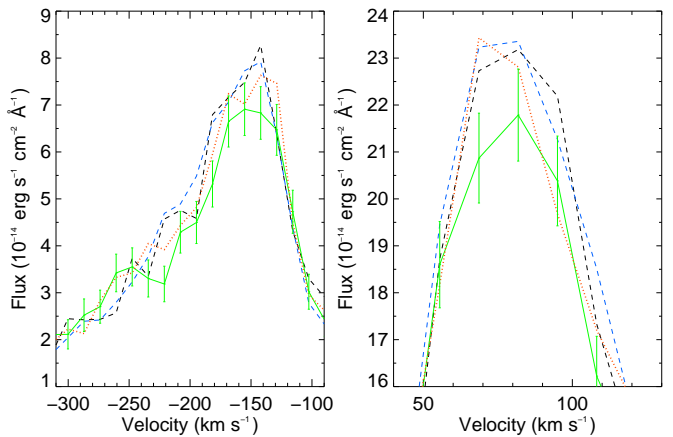}
\bigskip
\end{center}
\caption[]{Zoom on the blue wing (left panel) and red wing (right panel) of the Lyman-$\alpha$ line profile displayed in Fig.~\ref{Lyman_2011_4orbites}. The black and blue dashed lines show the fluxes before the planet transit; the green solid line with error bars at the 1\,$\sigma$ level shows the flux during the transit; the red dotted line shows the flux after the transit.} 
\label{Lyman_2011_4orbites_zoom}
\end{figure}

In contrast to the ACS spectra, the spectrally resolved STIS spectra allow us to look for absorption signatures within specific wavelength intervals (Fig.~\ref{Lyman_2011_4orbites}). 
Two clear absorption regions can be seen within the Lyman-$\alpha$ line in the 2011 transit observation. The most significant absorption signature is visible in the blue part of the line from \mbox{-230\,km\,s$^{-1}$} to -140\,km\,s$^{-1}$. Another feature, although less significant, is visible at the peak of the red wing from 60\,km\,s$^{-1}$ to 110\,km\,s$^{-1}$ (Fig.~\ref{Lyman_2011_4orbites_zoom}). There seems to be some absorption within the same velocity region in the blue wing for the post-transit observation, albeit with a lower depth. The red region displays no absorption feature in the post-transit observation. In the following sections we describe the methods used to quantitatively characterize these two signatures, in particular their precise velocity range and absorption depths.

\subsection{Quantitative estimates}
\label{best_range}

\begin{figure}[tbh]
\includegraphics[angle=-90,width=\columnwidth]{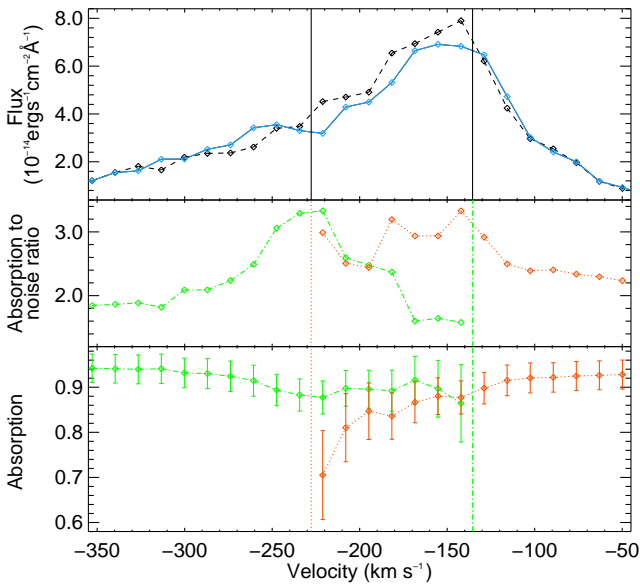}
\caption[]{Plot of the flux (upper panel), absorption (bottom panel), and ratio of the absorption to its noise (middle panel) in the blue part of the Lyman-$\alpha$ line, as a function of velocity. The upper panel displays the flux before the transit multiplied by the occultation factor of the planetary disk (black dashed line), to be compared to the flux during the transit (blue solid line). In the lower and middle panels, total cumulative values are calculated within domains of increasing range. The red dotted line is for ratios calculated within domains between \mbox{-230\,km\,s$^{-1}$} and increasing velocities. The green dash-dotted line is for ratios between -140\,km\,s$^{-1}$ and decreasing velocities.} 
\label{absblueshift}
\end{figure}

To quantitatively characterize the absorption features within the Lyman-$\alpha$ line, we studied the ratio of the flux during transit to the flux before transit. The transit signature is the combination of the absorption by the atmospheric neutral hydrogen and the 2.4\% occultation of the stellar disk by the planetary disk at all wavelengths. To characterize the absorption due to the hydrogen only, we decreased the flux before the transit by a factor corresponding to the planet absorption. Then, we searched for the strongest signature at every possible wavelength range, excluding intervals narrower than 0.1\,\AA\ (about two pixels in the STIS spectra). We looked for the most significant absorption, characterized by the highest value of the ratio of the relative absorption to its noise (Fig.~\ref{absblueshift}).

In the 2011 transit observations, the most significant signature is found within the range -230 to -140\,km\,s$^{-1}$, yielding an H\,{\sc i} absorption depth of 12.3$\pm$3.7\% (3.3\,$\sigma$ detection). This signature, due to the atmospheric hydrogen only, is the most significant over the entire Lyman-$\alpha$ profile. The measured velocity of the gas is higher than the escape velocity of 60\,km\,s$^{-1}$ from the planet (the effective escape velocity is even lower than 60\,km\,s$^{-1}$ at higher altitude where we observe the absorption). The total absorption depth of 14.4$\pm$3.6\%, which includes the planetary disk occultation, also corresponds to high altitudes: it is equivalent to a disk with a radius of 2.8 Jupiter radii. Therefore, the observed neutral hydrogen must be at high altitudes, escaping the planet gravity. Note that if we exclude the first HST orbit in the calculation of the flux before the transit, the absorption depth is the same within 1\,$\sigma$ (18.3$\pm$5.6\%), but the detected range is somewhat narrower, from -230 to -180\,km\,s$^{-1}$.

\begin{figure}[tbh]
\includegraphics[angle=-90,width=\columnwidth]{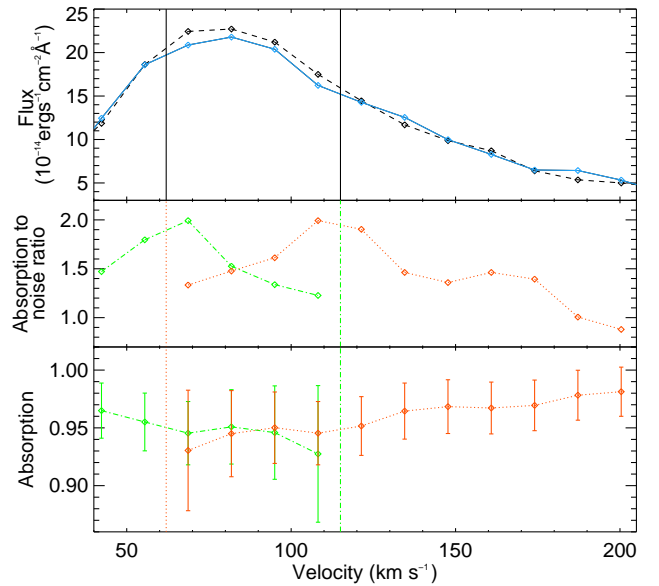}
\caption[]{Same as in Fig.~\ref{absblueshift} in the red part of the Lyman-$\alpha$ line. The red dotted line is for ratios calculated over domains between 60\,km\,s$^{-1}$ and increasing velocities. The green dash-dotted line is for ratios between 110\,km\,s$^{-1}$ and decreasing velocities.} 
\label{absredshift}
\end{figure}

	The same method was applied to the red wing of the Lyman-$\alpha$ line, with a search window in the range 40 to 200\,km\,s$^{-1}$. We find the most significant transit absorption signature in the range 60 to 110\,km\,s$^{-1}$  (see Fig.~\ref{absredshift}). Substracting the occultation by the planetary disk, this signature yields a 5.5$\pm$2.7\% absorption depth (2.0\,$\sigma$ detection).  
	This detection is the same if we exclude the first orbit, whether the spectra are corrected for the breathing effect or not. Cumulative transit depths due to the planetary disk and the atmospheric hydrogen can be found in Table~\ref{log_abs}. 

\begin{figure*}
\centering
\begin{minipage}[b]{.48\textwidth}			
\includegraphics[angle=-90,width=\columnwidth]{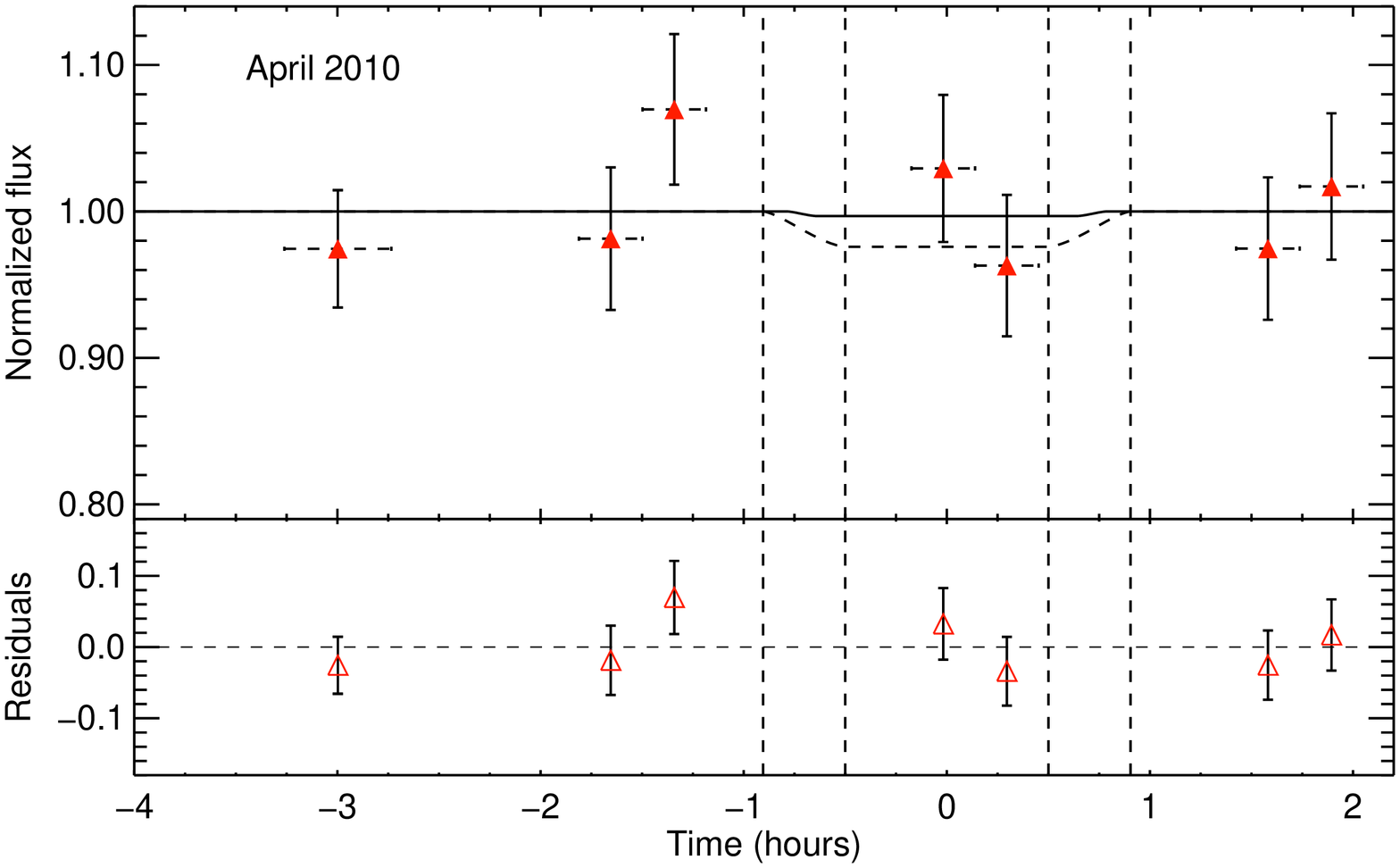}
\end{minipage}\qquad
\begin{minipage}[b]{.48\textwidth}			
\includegraphics[angle=-90,width=\columnwidth]{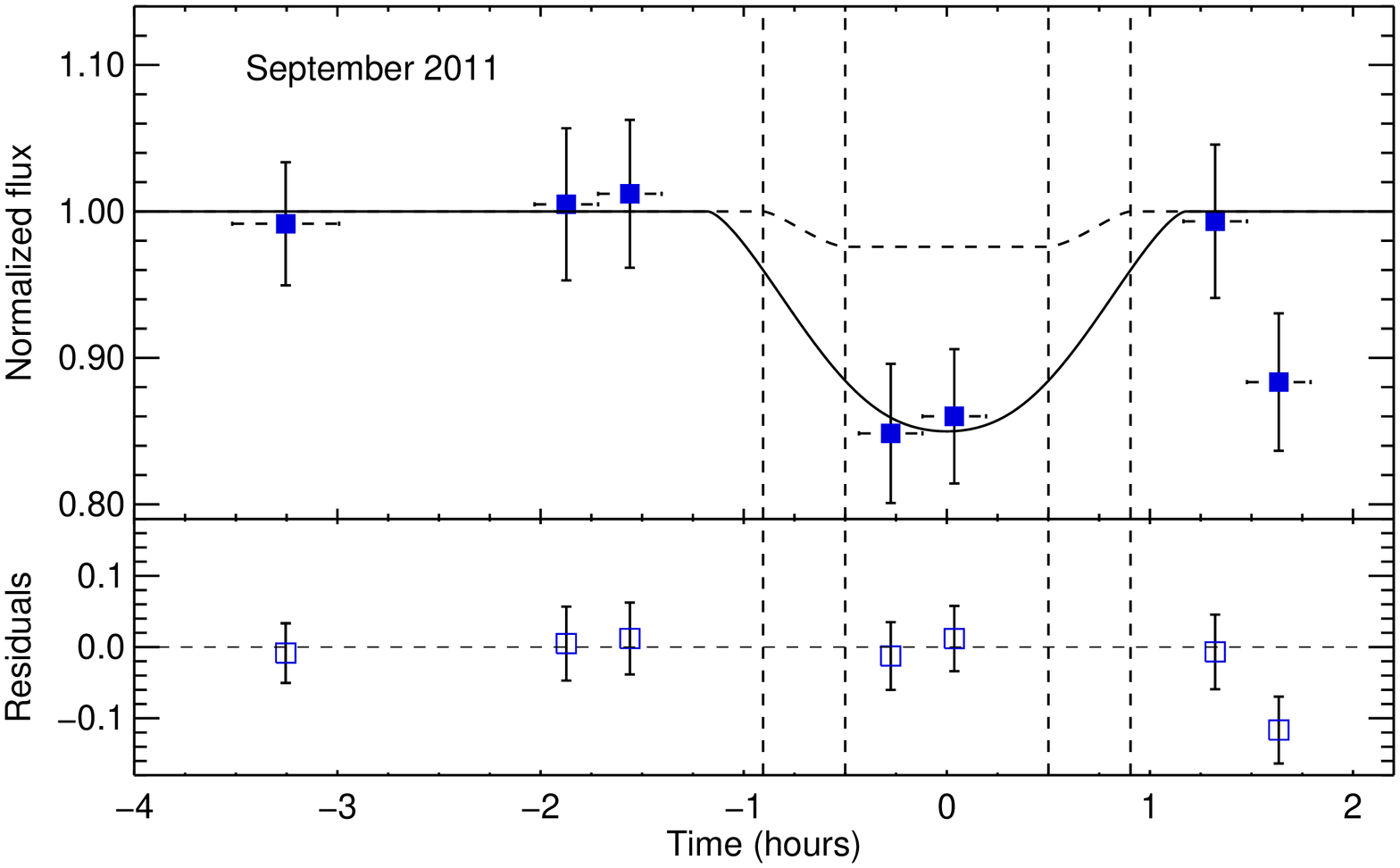}
\end{minipage}
\caption{Plot of the flux between -230 and -140\,km\,s$^{-1}$ in the blue wing of the Lyman-$\alpha$ line as a function of time relative to the center of the planetary transit (filled symbols). The red triangular symbols are for the 2010 observations, while the blue square symbols correspond to the observations of 2011. Each HST orbit is divided into two 1127~s independent subexposures, corrected for the breathing effect (except for the first orbit, see Section ~\ref{breathing}). Vertical dashed lines show the beginning and end of ingress and egress of the transit. Horizontal error bars centered on the symbols show the duration of the exposures. The light curve of the planet transit at optical wavelengths is displayed as a dashed black line. A light curve fitted with a classical planetary occultation model to the flux during the transit is displayed as a solid black line, and shows that the excess absorption feature detected in the 2011 data is not seen in the 2010 data. The bottom panel shows the corresponding residuals (empty symbols).}
\label{light_curves}
\end{figure*}

We also divided each exposure of a given HST orbit, corrected for the breathing effect, into two independent spectra with 1127~s exposure time. We measured the absorption over the range -230 to -140\,km\,s$^{-1}$ for the April 2010 and September 2011 observations. A light curve fitted to the data using a model of a single planetary disk occultation unambiguously shows that, in contrast to the 2010 measurements, the 2011 flux measurements require the presence of an extended
hydrogen atmosphere. Each of the two independent transit measurements of 2011 are consistent with a 14.4$\pm$3.6\% absorption, interpreted as atmospheric absorption in addition to the planetary disk occultation (Fig.~\ref{light_curves}). Note that the absorbing atmosphere is likely not spherical, which explains why the post-transit data of the 2011 observations cannot be fitted with the simple model used here (see Section \ref{post transit}). A full description of the geometry of the hydrogen cloud is beyond the scope of the present paper, however.

\begin{table*}[tbh]
\begin{tabular}{lccc}
\hline
\hline
\noalign{\smallskip}
Transit depths    				 & HST/ACS (2007 - 2008)		 & HST/STIS (2010) 				& HST/STIS (2011) 			\\
                					 & not spectrally resolved 	 & spectrally resolved	  & spectrally resolved	  \\
\noalign{\smallskip}
\hline
\noalign{\smallskip}
Planetary disk occultation & 2.4\%     						 		 & 2.4\%  						  	& 2.4\%    		    \\
Whole Lyman-$\alpha$ line  & 5.1$\pm$0.8\%         		 & 2.9$\pm$1.4\%  			  & 5.0$\pm$1.3\%   \\
-230 to -140\,km\,s$^{-1}$ & 											     & 0.5$\pm$3.8\%					& \textbf{14.4$\pm$3.6\%}  \\
60 to  110\,km\,s$^{-1}$ & 													 & 5.8$\pm$2.6\% 					& 7.7$\pm$2.7\%   \\
\noalign{\smallskip}
\hline
\hline
\end{tabular}
\caption{Transit depths in a given wavelength range (including planetary disk occultation and atmospheric absorption), for the three observations of HD\,189733b.}
\label{log_abs}
\end{table*}

\subsection{False-positive probabilities}
To assess the significance of the absorption features found in the Lyman-$\alpha$ light curve, we made the hypothesis that the transit spectrum shows no absorption and calculated with a bootstrap-like method the false-positive probability to have signatures as significant as the one we detected, but caused by noise only. 

From a reference-averaged spectrum, we generated two random spectra simulating spectra before and during the transit free of atmospheric absorption. The two spectra were obtained by adding random Gaussian noise to the reference spectrum, with an amplitude corresponding to the error estimated for the spectra before and during the transit. We generated these two spectra for 1,000,000 runs, and in each run used the absorption detection method detailed in Section~\ref{best_range} to find the most significant absorption signature. Due to a better consideration of the searched velocity intervals, we found lower estimates of the false-positive probability than in \citet{Lecav2012}. Indeed, taking into account that the range of the searched-for absorption feature cannot be narrower than the spectral resolution of the spectrograph, we excluded false-positive signatures found on wavelength intervals covering 1 or 2 pixels, which would be narrower than the STIS resolution. The false-positive probability to find an excess absorption depth at more than 3.3\,$\sigma$ in the range -350 to -50\,km\,s$^{-1}$ is only 3.6\%. It is unlikely that 
this particular signature comes from statistical noise in the data. The false-positive probability to find an excess absorption depth at more than 2.0\,$\sigma$ in the range 40 to 200\,km\,s$^{-1}$ is 24.6\%. With this higher probability, the 5.5$\pm$2.7\% absorption depth detected in the red wing is possibly due to noise in the data. Note, however, that a similar signature was also observed in the case of HD\,209458b, with an absorption depth of 5.2$\pm$1.0\% in the red wing between 1215.89 and 1216.43\,\AA~(\citealt{VM2003}, \citealt{VM2008}).

\subsection{Spectral and temporal variability}

\subsubsection{Absorption depth as a function of radial velocity}

\begin{figure}[tbh]
\includegraphics[angle=-90,width=\columnwidth]{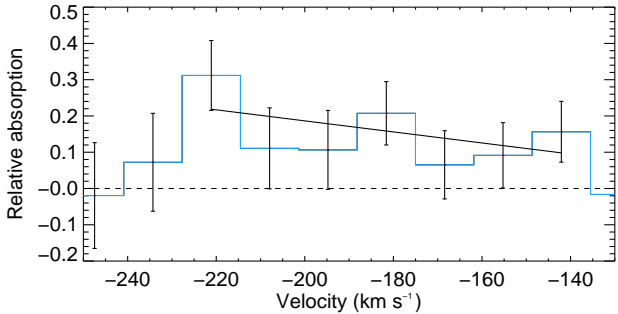}
\caption[]{Plot of the spectral absorption in the blue wing of the Lyman-$\alpha$ line in the 2011 observations (blue histogram). The black line is the best linear fit to the absorption and is not significantly different from a constant value.} 
\label{pente abs} 
\end{figure}

Here we address the possibility to resolve the absorption feature in the blue range of the Lyman-$\alpha$ line profile between -230 and -140\,km\,s$^{-1}$. Although the absorption depth seems to be higher for higher negative velocities, a linear fit to the spectral absorption within this domain shows that the slope is not significantly different from 0, with a value of (-1.5$\pm$1.3)$\times10^{-3}$\,(\,km/s)$^{-1}$ (Fig~\ref{pente abs}). We conclude that we do not see significant structures in the profile of the H\,{\sc i} absorption.

\subsubsection{Variations of absorption with time during the transit}
\label{abs with time}

\begin{figure}[tbh]
\includegraphics[angle=-90,width=\columnwidth]{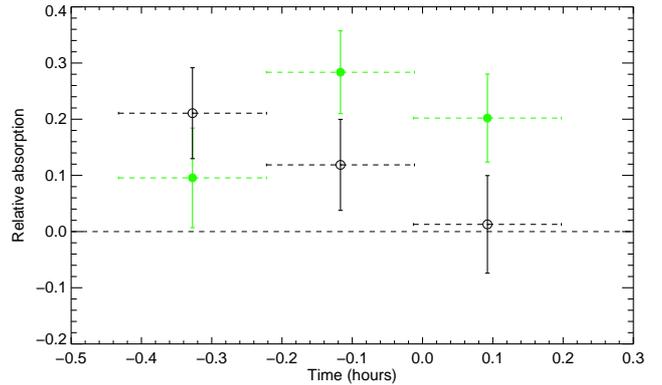}
\caption[]{Plot of the total absorption over two complementary velocity intervals, as a function of time during the transit, for three spectra with 751~s exposure time. Green disks stand for -230 to -175\,km\,s$^{-1}$ and black circles for -175 to -140\,km\,s$^{-1}$. Horizontal bars indicate the duration of each exposure.}
\label{379s abs} 
\end{figure}

The spectral resolution and high sensitivity of the STIS data allow the search for possible differences in the velocity structure of the absorption as a function of time \emph{during} the transit. Indeed, the escaping hydrogen may form a curved cometary tail trailing behind the planet (\citealt{Schneiter2007}; \citealt{Ehrenreich2008}). Since the tail crosses in front of the star, we can use time-tagged observations to probe different areas of the atmospheric cloud by measuring the absorption profile at different phases of the planet transit. We divided each exposure of a given HST orbit, corrected for the breathing effect, into three spectra with 751~s exposure time. We measured the absorption over the range -230 to -140\,km\,s$^{-1}$. The absorption over low negative velocities (\mbox{-175} to -140\,km\,s$^{-1}$) seems to decrease with time during the transit, while we see an increase of absorption for higher negative velocities (-230 to -175\,km\,s$^{-1}$) (see Fig.~\ref{379s abs}).

\begin{figure}[tbh]
\includegraphics[angle=-90,width=\columnwidth]{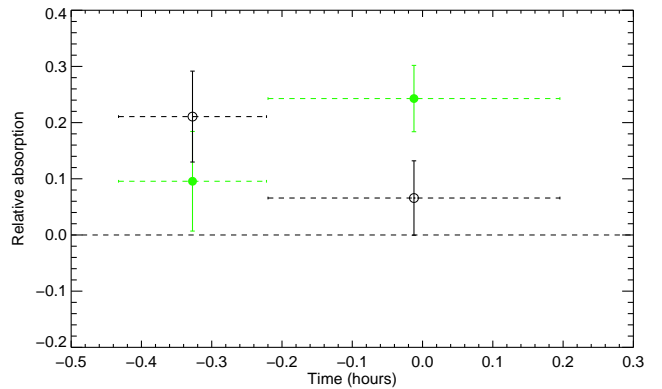}
\caption[]{Plot of the total absorption over two complementary velocity intervals, as a function of time during the transit, for two spectra with respectively 751~s and 1502~s exposure times. Green disks stand for -230 to -175\,km\,s$^{-1}$ and black circles for -175 to -140\,km\,s$^{-1}$. Horizontal bars indicate the duration of each exposure.}
\label{379s abs bis} 
\end{figure}

For the first of the three exposures (shortly after the end of the ingress) the difference between absorption at low velocities (21$\pm$8.1\%) and high velocities (9.6$\pm$8.9\%) is not significant. However, if we coadd the last two exposures to obtain the absorption at the center of the transit, we find that the absorption at low velocities is lower than for the first exposure, with a value of 6.6$\pm$6.6\%, while it has increased at high velocities, with a value of 24$\pm$5.9\% (Fig.~\ref{379s abs bis}). This is a 2\,$\sigma$ difference between the absorption depth at high and low velocities. Numerical simulations of the dynamics of the atmospheric cloud crossing in front of the star may help explain these variations (Bourrier et al. in prep). More observations of HD\,189733b transit will also be necessary to confirm possible variations of absorption with velocity over time.

\subsection{Post-transit observation}
\label{post transit}

\begin{figure}[tbh]
\includegraphics[angle=-90,width=\columnwidth]{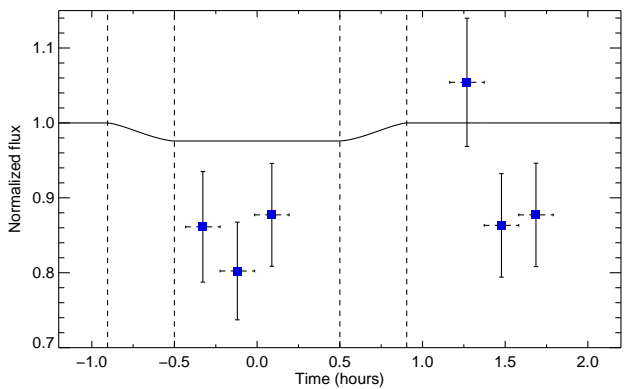}
\caption[]{Plot of the flux measured within the range -230 to -140\,km\,s$^{-1}$, for three 751\,s subexposures per orbit during the transit and after the transit, as a function of time (blue square symbols). Each flux, corrected for the breathing effect, is normalized by the flux of the corresponding exposure of the second orbit (i.e., the orbit just before the transit). Vertical dashed lines show the beginning and end of ingress and egress of the transit. Horizontal error bars centered on the symbols show the duration of the exposures in each HST orbit. The light curve of the planet transit at optical wavelengths is displayed as a solid black line.} 
\label{transit_3perorbit}
\end{figure}

Since the extended planetary atmosphere may form a curved cometary tail trailing behind the planet, the transit of the escaped hydrogen can last longer than the transit of the planet itself. Observations obtained after the transit may thus be affected by the signature of escaping atmospheric gas in the line of sight toward the star. This is the reason why, in the present analysis, we compared the Lyman-$\alpha$ line profile observed during the transit to profiles observed before the transit only, excluding post-transit observations. The best velocity interval \mbox{-230} to -140\,km\,s$^{-1}$, found for the in-transit observations, still yields an absorption depth of 6.83$\pm$3.82\% in the post-transit light curve (Fig~\ref{transit_3perorbit}).

Following the method detailed in Sect.~\ref{best_range}, we looked for the velocity interval providing the most significant absorption in the post-transit spectrum. We found the most significant absorption within the blue wing of the Lyman-$\alpha$ line in the range -230 to -180\,km\,s$^{-1}$, with a signature of 9.8$\pm$5.4\%. The higher velocities, with regard to the velocities of the transit signature detected in the transit spectrum, might be interpreted as absorption by atoms accelerated to their maximum velocity at the end of the cometary tail.

\section{Conclusions}

We observed the hot Jupiter HD\,189733b with the HST/STIS instrument in the far-UV wavelengths from 1195 to 1248\,\AA\, in April 2010 and September 2011. While no significant variation was detected in the N\,{\sc v} line at 1238.8\,\AA\ for both observations, the second epoch observation shows absorption in the Si\,{\sc iii} line and the N\,{\sc v} doublet at 1242.8\,\AA, albeit before the transit. This cannot be interpreted by a classical absorption in the planetary exosphere, and may be due either to an early occulation by a bow-shock or to intrinsic stellar line variations. 

We also presented a detailed analysis of the Lyman-$\alpha$ line of neutral hydrogen H\,{\sc i} observed in September 2011. A strong atmospheric absorption feature of 12.3$\pm$3.7\% (in addition to the 2.4\% absorption by the planetary disk) was found between -230 to -140\,km\,s$^{-1}$. The false-positive probability to find such a signature in the blue wing of the line was revisited to be only 3.6\%. Since HD\,189733 is an active star ({\it e.g.}, \citealt{Sanz-Forcada2011}), one might wonder if stellar activity might have been the source of this absorption signature. However, this feature is observed only within a very specific wavelength range, during the planet transit. The velocities of the absorbing hydrogen atoms, greater than the highest estimate of the escape velocity from the planet (60\,km\,s$^{-1}$) and explained by the interaction between stellar wind and hydrogen escaping the planet, unambiguously show that neutral hydrogen was escaping from HD\,189733b in 2011. 

HD\,189733b has been observed in the Lyman-$\alpha$ line during five transits at various epochs. The three transit spectra observed with the HST/ACS instrument in 2007/2008 (\citealt{Lecav2010}) displayed time variations in the transit depth at $\sim$2.5\,$\sigma$. The absorption depth measured in the HST/STIS spectra observed in September 2011 is consistent with the absorption depth in one of the three ACS spectra, and confirms the hydrogen escape from the planetary atmosphere. On the other hand, no absorption was found in the HST/STIS spectra observed in April 2010, which is consistent with the last of the three ACS spectra. Roughly half of HD\,189733b transit observations thus show detectable escape processes in neutral hydrogen.

Here we also report possible differences in the velocity structure of the absorption as a function of time \emph{during} the transit. The Lyman-$\alpha$ line also shows some absorption in the range -220 to -180\,km\,s$^{-1}$ in the post-transit spectrum. This is consistent with the idea of a cometary tail of accelerated hydrogen atoms trailing behind the planet. Upcoming observations of HD\,189733b in the Lyman-$\alpha$ line, along with a model of the hydrogen escape and its acceleration mechanisms, will help understand the temporal and spectral variability of the absorption in the exosphere of the planet.

\begin{acknowledgements}

Based on observations made with the NASA/ESA Hubble Space Telescope, 
obtained at the Space Telescope Science Institute, 
which is operated by the Association of Universities for Research in Astronomy, 
Inc., under NASA contract NAS 5-26555. We thank the referee for its clarifying comments.
D.E. acknowledges the funding from the European Commission's Seventh Framework Programme as a Marie Curie Intra-European Fellow (PIEF-GA-2011-298916).
G.E.B. acknowledges financial support by this program
through STScI grant HST-GO-11673.01-A to the University of Arizona.
These observations are associated with program \#11673. The authors acknowledge financial support
from  the Centre National d'\'Etudes Spatiales (CNES). This work has also been supported
by an award from the \textit{Fondation Simone} et \textit{Cino Del Duca}.

\end{acknowledgements}

\bibliographystyle{aa} 
\bibliography{biblio} 

\end{document}